\documentclass[%
preprint,
 amsmath,amssymb,
aps,
]{revtex4-2}

\usepackage{graphicx}
\usepackage{dcolumn}
\usepackage{bm}
\usepackage{textgreek}
\usepackage[dvipsnames]{xcolor}

\newcommand{\anais}[1]{\textcolor{black}{#1}}

\begin{document}


\title{A new pressure sensor array for local normal stress measurement in complex fluids}

\author{Anaïs Gauthier*, Mickaël Pruvost, Olivier Gamache and Annie Colin}
\email{anais.gauthier@espci.fr or annie.colin@espci.fr}
\affiliation{
MIE – Chemistry, Biology and Innovation (CBI) UMR8231, ESPCI Paris, CNRS, PSL Research University, 10 rue Vauquelin, Paris, France}


\vspace{3 cm}
\begin{abstract}
\anais{A new pressure sensor array, positioned on the bottom plate of a standard torsional rheometer is presented. It is built from a unique piezo-capacitive polymeric foam, and consists of} twenty-five capacitive pressure sensors (of surface 4.5$\times$4.5 mm$^2$ each) built together in a 5$\times$5 regular array. The sensor array is used to obtain a local mapping of the normal stresses in complex fluids, which dramatically extends the capability of the rheometer. \anais{We demonstrate this with three examples. First, the pressure profile is reconstructed in a polymer solution, which} enable the simultaneous measurement of the first and the second normal stress differences $N_1$ and $N_2$, \anais{with a precision of 2 Pa. In a second part, we show that negative normal stresses can also be detected. Finally, we focus on the normal stress fluctuations that extend} both spatially and temporally in a shear-thickening suspension of cornstarch particles. \anais{We evidence the presence of local a unique heterogeneity rotating very regularly.} In addition to their low-cost and high versatility, the sensors show here their potential to finely characterize the normal stresses in viscosimetric flows.
\end{abstract}



\maketitle

\newpage

\section{\label{intro}Introduction} 
Contrary to Newtonian fluids which are fully characterized by their viscosity $\eta$ only, complex fluids exhibit a much diverse behavior. Their micro-structure, at an intermediate length scale between the molecule and the sample size (the radius of gyration of polymer coils in polymer solutions or the size of solid particles in suspensions) is altered by the flow. This often induces a non-linear relation between the applied stress $\sigma$ and the resulting flow velocity: its understanding is a fundamental challenge, central in many industrial processes. A striking example is the presence of normal stresses, appearing in the diagonal components of the stress tensor. In polymer melts and solutions, the normal stresses are responsible for a number of spectacular effects, such as the swelling of the fluid in extrusion processes, or its climbing on rotating rods \cite{Bird:1987}, and have been the object of numerous studies \cite{Adams:1964, Keentok:1980, Gao:1981, Meissner:1989, Schweizer:2002, Alcoutlabi:2009}. Normal stresses are much less documented in other systems, even if they are essential to fully characterize the flow of complex materials. Yield-stress fluids, for example develop moderate normal stresses in shear, which sign (positive or negative) and origin are still actively discussed \cite{Montesi:2004, Seth:2011, deCagny:2019}. In suspensions of solid particles, \anais{the normal stresses are still being characterized \cite{Zarraga:2000, Couturier:2011, Dbouk:2013, Mari:2014, Lobry:2019}. They are causing, in particular, particle migration \cite{Boyer:2011} or edge fracture \cite{Keentok:1980, Tanner:2019}. The normal stresses also increase dramatically in dense shear thickening suspensions:} very high normal stresses, up to 10 000 Pa have been reported \cite{Fall:2008,Ovarlez:2020}. They are often associated with with an inhomogeneous flow, and generate localized forces strong enough to damage the rotors of mixing systems \cite{Ovarlez:2020}.

\noindent In viscosimetric flows, normal stresses are far less studied than shear stresses, mostly because their measurement is not straightforward. \anais{Indeed, on} a torsional rheometer, the normal stress differences $N_1$ and $N_2$ are usually measured by combining two experiments with different geometries: the net thrust force gives access to $N_1$ in a cone/plate geometry, and to $N_1-N_2$ in a parallel plate geometry \cite{Bird:1987, Ginn:1969, Dai:2013, Tanner:2016, deCagny:2019}. $N_2$ is then calculated by finding the often small difference between two large experimental quantities, which amplifies the impact of any measurement error \cite{Alcoutlabi:2009}. Other techniques have been developed to obtain a more reliable measurement of $N_2$ \anais{\cite{Tanner:1970}}, or to avoid using a cone plate geometry (unsuitable in some systems such as suspensions of large particles): the rotating rod rheometry \cite{Beavers:1975, Magda:1991, Zarraga:2000, Boyer:2011}, \anais{measurement of the shape of a free surface \cite{Keentok:1980}} or the tilted through method \cite{Tanner:1970, Couturier:2011}. Another method consists of measuring the pressure distribution as a function of the radial position $r$ in a cone/plate or a plate/plate geometry, by using a small number of pressure sensors integrated to the plate of a torsional rheometer. \anais{This technique, which gives both $N_1$ and $N_2$ in a single experiment, has been attempted in polymer solutions \cite{Adams:1964, Miller:1972, Gao:1981, Alcoutlabi:2009} and recently in a non-Brownian suspension \cite{Dbouk:2013}.}\\
\noindent \anais{Unfortunately, such setups are complex to build and measurements can be flawed. For example, Couturier \textit{et al.} \cite{Couturier:2011} use a rectangular channel that causes unwanted secondary flows near the corners of the cross-section.} The use of sensors must also meet a number of important requirements, for example being small enough relatively to the rheometer plate, be extremely sensitive and not cause any disturbance to the local flow \cite{Baek:2003}. \anais{The pressure transducer membranes have to be positioned at exactly the same level as the disk surface: if not, a hole pressure has to be accounted for \cite{Tanner:1969, Pritchard:1970}, and may lead to error measurements, as in the pioneering work of Adams and Lodge \cite{Adams:1964}. To this purpose, Dbouk and coworkers \cite{Dbouk:2013} coated the surface transducers with paraffin, so that no pressure hole effect is expected to take place. In addition, the number of sensors is often limited by the size of the plate and by the volume of the acquisition system.} \\
\noindent Here, we demonstrate the potential of a new low-pressure sensor array in the measurement of local normal stresses in non-Newtonian fluids. The number of sensors (25 in a 4$\times$4 cm$^2$ surface), their precision (up to 2 Pa) and the frequency of measurements (200 Hz) are unprecedented. In addition, the sensor array is highly versatile: the position, the size and the number of sensors is easily varied by changing only the bottom electrode. \anais{They also exhibit no drift and involve no pressure hole as the contact surface is flat and made of one piece.} \anais{After presenting the sensor fabrication method and acquisition system, we demonstrate its potential to characterize three different flows, in a parallel plate geometry. In a polymer solution, we measure simultaneously the two normal stresses $N_1$ and $N_2$, both in very concentrated and very diluted systems. We then focus on a Newtonian fluid subjected to a secondary flow, and measure this time negative normal stresses. Finally, we consider a shear-thickening fluid submitted to flow heterogeneities: our sensors evidence the presence of a single aggregate rotating very regularly, an intriguing phenomenon that cannot be detected through the force sensor of the rheometer.}

\anais{\section{\label{sec1}The sensor array: fabrication, calibration and properties}}

\begin{figure}[!ht]
\centering
\includegraphics[width=0.99\columnwidth]{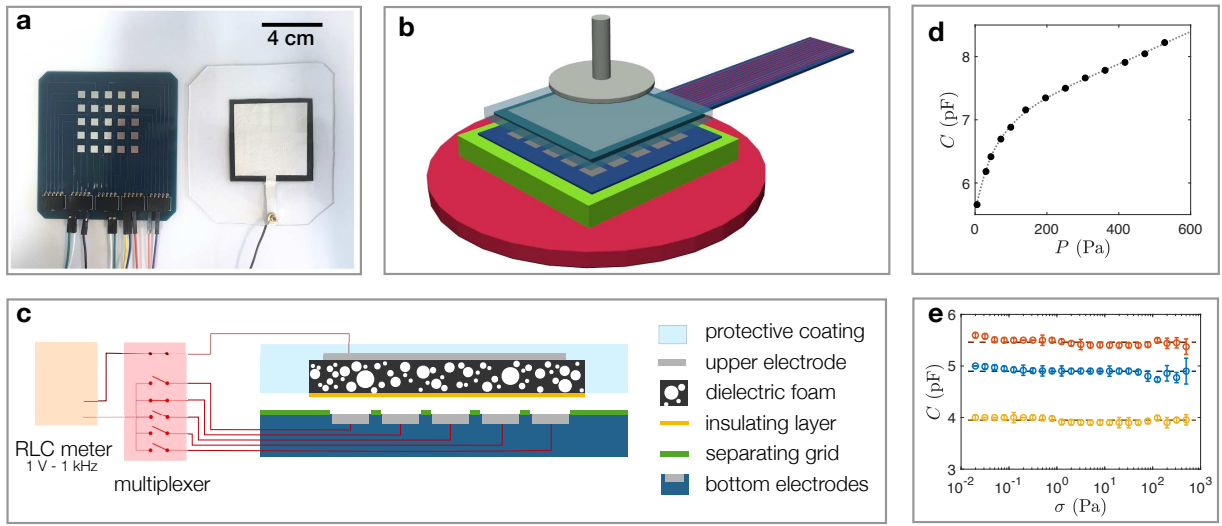}
\caption{\label{figure1}\textbf{a.} Picture of the electrode network (left) and the measurement surface (right) of a sensor array. \textbf{b}. Experimental system: both layers are placed on top of each other on the bottom plate of a parallel plate rheometer. The upper disk has a radius $R$ of 2 cm, and the gap thickness $h$ is 1 mm. \textbf{c}. Side-cut of a capacitive sensor array. \textbf{d.} Typical calibration curve of one sensor for applied pressures $P>0$, showing the capacitance $C$ as a function of a hydrostatic pressure $P$. \anais{\textbf{e.} Capacitance of three sensors when subjected to varying shear stress $\sigma$. Blue dots correspond to a central sensor (at a position $r$ = 0 to the center of the geometry) while the red and yellow dots correspond to two sensors positioned diagonally at $r$ = 12 mm.}}
\end{figure}

\noindent \textbf{Sensor array.} \anais{A distinctive feature} of the pressure sensor array \anais{presented} here is that the 25 sensors are not designed individually. As shown in figure \ref{figure1}a, the sensor array consists of two surfaces:  \anais{a solid \textit{electrode network} (left) and a soft \textit{measurement surface} (right) made of a polymeric material. As shown in Figure \ref{figure1}b, the electrode network is placed on the bottom plate of a torsional rheometer (Discovery HR-2, TA instruments) and covered by the measurement surface. A thin grid of double sided tape is added between the two surfaces. The 3-layer sandwich thus formed is the pressure sensor array itself, presented in sectional view in Figure \ref{figure1}c.} 

\noindent \anais{The bottom layer is shown in Figure \ref{figure1}c, in dark blue and gray. It is a simple electrode network, custom-made by JLC PCB. The electrodes are 25 conductive square surfaces of 4.5$\times$4.5 mm$^2$,  organised in a 5 by 5 regular matrix, and connected individually to the acquisition system. The total surface of the network is 4$\times$4 cm$^2$, which corresponds to the typical size of a rheometer plate geometry.} 

\noindent\anais{On top of the electrode network is attached a 25 \textmugreek m thick layer of double-sided tape, shown in green in Figure \ref{figure1}c. Using a laser cutter, 25 square holes (with size 5$\times$5 mm$^2$) are cut into the tape so that it does not cover the electrodes. The role of this intermediate layer is double: first, it ensures a good adhesion of the bottom layer to the measurement surface. In addition, its presence increases the sensitivity of the sensor by a factor 10 at very low pressures ($<$ 150 Pa). We interpret this as the consequence of a local bending of the soft measurement surface within the grid (by typically 10 to 20 \textmugreek m) which can only happen when the measurement surface is slightly raised above the bottom electrodes.} 

\noindent \anais{The last layer is the measurement surface itself. Its core is a piezo-capacitive soft solid foam (shown in black in Figure \ref{figure1}c), made from a polydimethilsiloxane polymer (PDMS) filled with 10\% in weight of carbon black particle. This material was developed previously in our team, and its fabrication method is presented in detail in Refs. \cite{Pruvost:2018, Pruvost:2019}. Here, the solid foam is fully integrated into a multi-layered sandwich to make a solid and reliable sensor. 
The foam itself is prepared as described in \cite{Pruvost:2019}, through a water-in-oil emulsion process: a mixture of water and carbon-black particles is slowly added to PDMS, under vigorous mixing. The emulsion then forms a paste, that is uniformly spread in a 1 mm thick sheet using an applicator (Zehntner ZUA 2000). The spreading is done on a Mylar surface covered by an 5 \textmugreek m thick insulating layer of plain PDMS (shown in yellow in Figure \ref{figure2}c), previously deposited with a spin-coater and cured for 1h at 70$^{\circ}$C. After spreading, the paste is cured in a two-step process. First, it is placed in a bath of deionized water at 70$^\circ$C for 6 h to ensure cross-linking of the PDMS polymer. After curing, the solid is let to dry at 70$^{\circ}$C for 24 h so that all water contained in the pores evaporates. After curing and drying, the emulsion turns into a piezo-capacitive soft solid foam. Its relative softness (with a Young modulus of 1.6 MPa \cite{Pruvost:2018}) is at the source of its remarkable piezo-capacitive properties: when a pressure is applied on the material, the micropores (of size between 1 and 10 \textmugreek m) deform, which induce a large variation of its permittivity and thus its local capacitance. In the sensor array, the upper part of the foam is covered by a 25 \textmugreek m thick paste of silver particles (Creative Materials), shown in gray in Figure \ref{figure1}c, which plays the role of a soft electrode. The upper electrode is connected to the acquisition system by planting a conductive screw attached to a wire through in the silver film (as in Figure \ref{figure1}a). Finally, a protective coating of PDMS is cast around the three layers of the measurement surface. The PDMS, initially liquid, is let to rest for 30 min at ambient temperature to ensure that the measurement surface is flat. It is then cured at 70$^\circ$C for 1h. We thus obtain a good alignment of the sensor with the upper plate geometry. Small fluctuations of the thickness (of the order of 20 \textmugreek m) are observed when determining the zero position of the upper plate. These perturbations are small compared with the gap size in our experiments (1 mm) and the radius of the geometry (2 cm) and they are considered as standard in rheology.}

\noindent \anais{The Young modulus of the whole measurement surface is close to the modulus of the piezo-capacitive foam alone ($\simeq$ 1.6 MPa \cite{Pruvost:2018}). At small pressures (of the order of 100 Pa), the main source of deformation of the sensor originates in its local bending in the grid of double-sided tape. The level difference between the disc and the coated sensor is thus less than 25 \textmugreek m, while the gap
between the discs is set to 1 mm. In such conditions, no pressure hole effect is expected.
To check if the softness of the sensor and the perturbations of the measurement surface might impact the flow, we compare in Supplementary Figure 1 the flow curves of the two non-Newtonian fluids studied here (a polymer solution and a cornstarch suspension) when measured directly on the (solid) rheometer plate or on the soft sensor array. In both cases, the flow curves are almost identical, which indicates that the presence of the sensor does not significantly disturb the flow.\\}


\smallskip

\anais{\noindent \textbf{Acquisition system.} Due to its softness, the measurement surface does not redistribute the pressure and the three-layers sandwich sensor array behave as 25 independent piezo-capacitive elements, with a capacitance $C$ directly correlated to the local pressure. Each sensor capacitance $C$ is recorded as a function of time by an acquisition system, schematized in Figure \ref{figure1}c. It consists of two instruments: a multiplexer (Keysight 34980A Agilent technologies) and a precision LCR meter (Keysight E4980AL Agilent technologies), both controlled using a in-house Matlab code. The LCR meter imposes a sinusoidal signal at low frequency (1 kHz) and low voltage (1 V), and measures the capacitance of the connected circuit with a precision of 0.05 \%. Its internal impedance is automatically adjusted to the circuit: here it is set to 5 k$\Omega$. The multiplexer consists of 25 optical switches (one per sensor), with response time 0.2 ms. During a measurement, the following steps are repeated: first, the switch corresponding to one bottom electrode is closed, and the LCR meter records the capacitance of the circuit between this electrode and the top electrode, averaged over 50 ms. The switch is then closed. This protocol is repeated for every sensor of the array -- which takes approximately 1.5 s for all 25 sensors -- before the next measurement cycle starts. In the following analysis of the data, an internal compensation calculation is done to account for the impedance of the multiplexer and wires. In most experiments, we use between 5 and 10 sensors, which automatically increases the measurement frequency, up to 2-3 cycles per second.}

\noindent \anais{For higher frequency measurements, we use an electronic circuit custom made by Piwio, which miniaturizes the acquisition system. The circuit is similar in principle to the one described above, with three main differences. First, square waves of 1 V amplitude are used, with frequency 1 kHz. Second, there is one multiplexer for every 4 sensors, which increases the measurement frequency. Finally, the capacitance is measured with a low resolution of 0.1 pF, and it is not averaged. This system is much faster than the previous one (up to 200 Hz), but less accurate. In Supplementary Figure 2, the typical response of the sensors when connected to the two acquisition systems are compared. When using the precision LCR meter, the noise (of the order of 1.5 Pa) is higher than the resolution of the acquisition system (0.00001 pF, which corresponds to a pressure of $10^{-3}$ Pa). With this measurement method, pressures as low as 2 Pa can be measured, as further confirmed in Figure \ref{figure3}c. The Piwio integrated board, on the other hand, is limited by its resolution and cannot measure pressures below 10 Pa. These values of 2 and 10 Pa, which set experimentally the pressure detection threshold of our sensors, are close to the 4 Pa threshold of our rheometer (Discovery HR-2, TA instruments). However, since the surface of the sensors is 65 times smaller than the surface of the rheometer geometry, the sensor array is sensitive to forces typically 50 times smaller than the rheometer.} 

\smallskip

\textbf{Sensor calibration.} \anais{The sensors are calibrated using hydrostatic pressure in compression, and using a well-known flow in extension.} In compression, a cylinder (with diameter 6 cm) is placed on top of the measurement surface and made water-tight with grease. Water is then added, and the capacitance $C_i$ of each sensor is measured as a function of the water pressure $P = \rho g h$ (with $h$ the liquid height, $\rho$ water density and $g$ gravity). A calibration curve is presented in Figure \ref{figure1}d: the sensor capacitance $C$, initially equal to a $C_0 =$ 5.5 pF, increases by 50\% for a pressure variation $\Delta P$ of 600 Pa. The sensitivity $S = \frac{\Delta C/C_0}{\Delta P}$ is higher at low pressures, with $S$ = 1.8 kPa$^{-1}$ for $P<$ 100 Pa. It remains excellent at higher pressures, with $S = $ 0.7 kPa$^{-1}$ for $P >$ 200 Pa. \anais{While the sensors are particularly sensitive to the applied pressure, they do not react to shear stress. This is shown in Figure \ref{figure1}e, where the sensors capacitance is recorded when shearing silicone oil with viscosity 100 000 cSt. At the stresses $\sigma$ considered here, silicone oil is a Newtonian fluid, with a constant viscosity and no measurable normal stress (see Supplementary Figure 3). The response of 3 sensors, placed either at the center of the geometry (blue dots) or at an equal distance $r$ = 12 mm to the center (red and yellow dots) is shown. While the applied shear stress $\sigma$ is varied over 5 orders of magnitude, the sensors capacitance remains constant.}
\noindent \anais{Finally, the sensors are calibrated in extension using a Newtonian fluid sheared at high velocity. Indeed, in a parallel plate geometry, a recirculation appears, which generates a negative pressure close to the static plate, varying quadratically with the angular velocity $\Omega$ of the rotating plate (see section \ref{sec3} for more details). We use this theoretical pressure value as a reference to calibrate our sensors.\\}

\noindent \anais{Two other properties of the sensor are also considered. First, we estimated the sensor response time by applying a constant pressure load (of a few hundred of Pa) on the sensors and removing it suddenly. The sensors response time is almost instantaneous to increasing pressures (with a response time of the order of 50 ms), and slightly longer to decreasing pressures (100 to 200 ms). In the experiments considered here, the measurement frequency typically varies between 1 and 20 Hz: we can thus consider that the sensor response is instantaneous, even in dynamic experiments. We finally checked that the sensors do not exhibit any significant drift in time. This is demonstrated in Supplementary Figure 4a: the response of the sensors, submitted to a constant hydrostatic pressure of 300 Pa is shown to remain constant. In addition, we compare in Supplementary Figure 4b the sensors response before and after a 2000 s experiment, after being subjected to pressures varying between 100 and 400 Pa. The difference of measured pressure at zero angular velocity before and after 2000 s is smaller than 4 Pa, which is of the order of the accuracy of the sensors. The potential drift is thus extremely small compared to what is reported in the literature \cite{Dbouk:2013, deCagny:2019}, which can reach 15 to 30 Pa in 60s \cite{deCagny:2019}}

\medskip

\noindent \textbf{Rheology procedure.} We use a parallel plate geometry, with the measurement surface of the sensor as a static bottom plate. Liquid is added directly on the measurement surface. After positioning the upper disk at the working gap ($h$ = 1 mm) the liquid in excess is carefully cleaned, so that the meniscus at the edge of the plate \anais{is as vertical as possible}. At this stage, both the internal force sensor of the rheometer and the sensor array are set to 0. This consists in assuming that the system is pre-stressed by a constant capillary pressure $P_c$ (typically of the order of 50 Pa) due to the meniscus, that is subtracted in the pressure measurements. This assumption of a constant \anais{pressure at the edge is valid as long as instabilities such as edge fracture \cite{Keentok:1980, Tanner:2019} can be avoided. Such instabilities are expected to appear in fluids with a negative second normal stress $N_2$, when $\lvert N_2\rvert$ exceeds a critical value $N_c \simeq 5 \gamma/h$, with $\gamma$ the surface tension of the fluid and $h$ the gap size. In our experiments with a polymer solution of HPAM, $N_2$ never exceeds 16 Pa, which is much smaller than $N_c \simeq 350$ Pa. In the particle suspension system, edge effects do not directly impact our results, focused on the detection of flow heterogeneities.}

 \noindent During an experiment, a constant shear stress $\sigma$ = $\sigma_{\theta z}$ is imposed on the fluid by the upper disk (with radius $R$ = 2 cm). The average normal stress is obtained with the built-in force sensor of the rheometer, which measures the mean vertical thrust on the geometry. \anais{The sensor array, on the other hand measures the local pressure $P(r,\theta) = -\sigma_{zz}$ within the sheared fluid, with $\sigma_{zz}(r,\theta)$ the $zz$ component of the stress deviator tensor in the position ($r,\theta,z = 0$).}

\smallskip

\noindent In the limit of low Reynolds numbers, the velocity $\mathbf{v}$ of the fluid is (almost) purely orthoradial, with $\mathbf{v} =  v_{\theta}\,\mathbf{e}_{\theta} = r\frac{\Omega z}{h}\,\mathbf{e}_{\theta}$ and $v_r, v_z \ll R \Omega$. The shear rate $\dot{\gamma}$ thus writes: $\dot{\gamma} = \frac{r\Omega}{h}$. For an homogeneous liquid, the equations of motion write, in the $r$ direction:

\begin{equation}
    \label{motion}
    \frac{1}{r}\frac{\partial}{\partial r}(r\sigma_{rr}) - \frac{\sigma_{\theta \theta}}{r} = -\rho \frac{v_{\theta}^2}{r} + \frac{\partial p}{\partial r}
\end{equation}

\noindent The diagonal components in the stress tensor $\sigma$ thus generate a radial variation of the pressure. This will be illustrated in the next section, using an emblematic viscoelastic fluid \anais{known to generate significant normal stresses.}

\section{\label{sec2} Positive normal stress measurements}

Long-chain polymer solutions are viscoelastic fluids, which flow is characterized by three functions: their viscosity $\eta = \sigma/\dot{\gamma}$, the first normal stress difference $N_1 = \sigma_{\theta \theta} - \sigma_{zz}$ and the second normal stress difference $N_2 = \sigma_{zz} - \sigma_{rr}$. In polymer melts and concentrated polymer solutions, $N_1$ is typically positive and high, while $N_2$ is more than 10 times smaller and negative \cite{Bird:1987}. 

\begin{figure}[!ht]
\centering
\includegraphics[width=0.999\columnwidth]{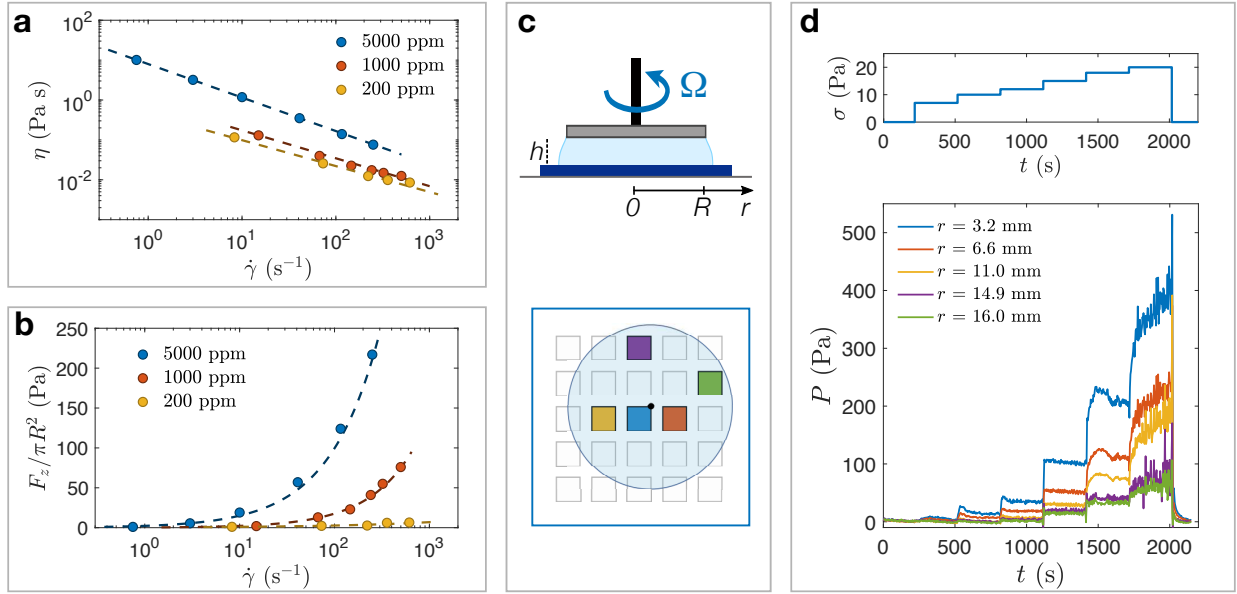}
\caption{\label{figure2} Rheological measurements of HPAM solutions
\textbf{a.} Viscosity $\eta$ as a function of the shear rate $\dot{\gamma}$ for three different concentrations of a HPAM solutions: 5000 ppm (blue), 1000 ppm (red) and 200 ppm (yellow). \textbf{b.} Average vertical pressure $F_z/\pi R^2$ measured through the built-in sensor of the rheometer, as a function of $\dot{\gamma}$. \textbf{c.} (top) Side view of an experiment. The local pressure profile $P(r)$ is measured as a function of the radial distance $r$ to the center of the geometry. (bottom) Top view of the sensor array, with the sensors used in the experiment shown in bright colors. \textbf{d.} Response of the sensors during a typical experiment of stress-imposed steps $\sigma(t)$. The pressure $P = -\sigma_{zz}$ measured by 5 different sensors is plotted as a function of time $t$.}
\end{figure}

\noindent Using the pressure sensor array, both $N_1$ and $N_2$ are measured in a single experiment. We use a solution of a partially hydrolyzed polyacrylamid polymer (HPAM) with a high molar mass $M_w = 18.10^6$ g/mol. HPAM is a charged linear polymer consisting of acrylamid monomers, where 25 to 30\% of the amine group are replaced by a carboxyl group. The concentration of HPAM in water is varied from 200 ppm (very diluted) to 5000 ppm (very concentrated). As shown in Figure \ref{figure2}a, the rheology of these solutions is characteristic of a viscoelastic fluid, with a strong shear thinning: the viscosity $\eta$ decreases with $\dot{\gamma}$ following a power law $\eta \propto |\dot{\gamma}|^n$. $n$ increases in magnitude with the concentration: on Figure \ref{figure2}a, the best fit (dotted lines) is obtained for $n= -0.84$ for 5000 ppm, $n = -0.70$ for 1000 ppm and $n = -0.65$ for 200 ppm. At high shear rates, the liquid develops normal stresses, which average value is measured by the build-in force sensor of the rheometer, and presented in  Figure \ref{figure2}b. The thrust $F_z/\pi R^2$ averaged on the plate surface is plotted as a function of $\dot{\gamma}$ for three HPAM solutions. The concentrated solutions exhibit high normal stresses: for example, \anais{the mean normal stress of the 5000 ppm} solution is more than ten times higher than the shear stress $\sigma$ at high shear. In the region where they are present, \anais{the normal stresses} increase proportionally to $\dot{\gamma}^m$. $m$ is only measurable for the most concentrated solutions: we find $m$ = 0.8 for the 5000 ppm solution and $m$ = 0.9 for the 1000 ppm solution. 

\noindent \anais{The sensors measure the pressure $P = -\sigma_{zz}$} as a function of the radial distance $r$ to the center of the geometry, as defined in Figure \ref{figure2}c (top). \anais{For simplicity, we present here the pressure measurement of only 5 different sensors placed at varying distances $r$ to the center of the geometry, and distributed in all directions. Indeed, as presented in Supplementary Figure 5, two sensors placed at an identical distance $r$ to the center of the geometry measure the same pressure.} On figure \ref{figure2}c, the  sensors are colored in blue ($r$ = 3.2 mm), red ($r$ = 6.6 mm), yellow ($r$ = 11.0 mm), purple ($r$ = 14.9 mm) and green ($r$ = 16.0 mm). The geometry is shown in light blue: its center (indicated by a black dot) is placed at the level of the upper right corner of the central sensor. 

\noindent A typical experiment is presented in Figure \ref{figure2}d, with a 5000 ppm HPAM solution. The shear stress $\sigma$ (plotted on top) is increased by regular steps of 300 s, from $\sigma = 0$ Pa to $\sigma$ = 20 Pa. \anais{The stress $P = -\sigma_{zz}$ measured by each sensor is recorded as a function of time $t$}, and presented in Figure \ref{figure2}d with the same color code as in Figure \ref{figure2}c. As visible in Figure \ref{figure2}d, $P$ increases with $\sigma$. At each stress-imposed step, the higher pressures are measured close to the center of the geometry (in blue): the pressure continuously decreases with $r$. \anais{While it remains almost constant for $\sigma \leq 15$ Pa (with a very slight decrease in time), the pressure signal becomes noisy for $\sigma > 15$ Pa. We interpret this phenomenon as a consequence of an elastic turbulence phenomenon, which are likely to occur at low Reynolds numbers in our systems \cite{Kawale:2017, Bodiguel:2015, Yao:2020}. Such instabilities are driven by the stretching of the polymer chains in the azimutal direction. Following Pakdel and Mc Kinley \cite{Pakdel:1996, Mckinley:1996}, we expect these instabilities to appear in the a parallel plate geometry when the criterion $\left( h/R \right)^{1/2} N_1/\sigma > M$ is met, with $R$ the streamline curvature and $M$ a constant, ranging between 1 and 6 depending on the rheological law of the fluid \cite{Larson:1994, Muller:2008}. Anticipating that $\lvert N_2 \rvert \ll N_1$, we calculate the Weissenberg number at the transition as Wi = $N_1/\sigma \simeq F_z/(\pi R^2 \sigma) \simeq 10$, which leads to $M \simeq 3$. This value is in good agreement with what is expected for elastic turbulence \cite{Yao:2020}.}

\smallskip

\begin{figure}[!ht]
\centering
\includegraphics[width=0.7\columnwidth]{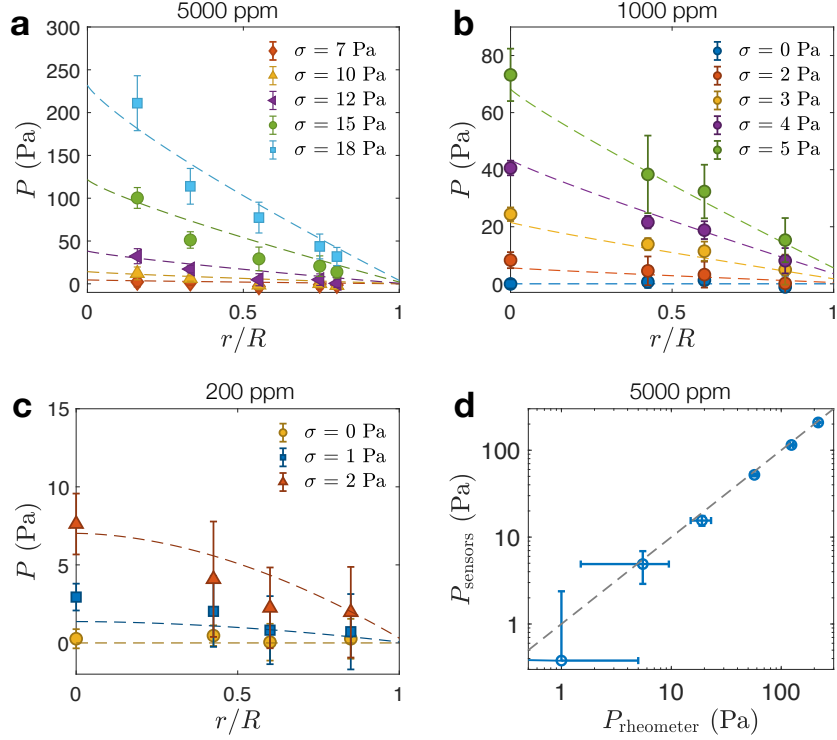}
\caption{\label{figure3}Pressure profiles $P(r)$ within flowing HPAM solutions in a plate/plate configurations, for varying shear stresses $\sigma$. The dotted lines are the best fit of the model, with $P(r) = - \dot{\gamma_R}^m\left[ \frac{\alpha_1 + (m+1)\alpha_2}{m}\left(\frac{r}{R}\right)^m - \frac{\alpha_1 + \alpha_2}{m} \right]$. \textbf{a.} Solution of 5000 ppm HPAM in water: $\alpha_1$ = 4.7, $\alpha_2 = -0.2$. \textbf{b.} HPAM 1000 ppm: $\alpha_1 = 0.46,\,\alpha_2 = -0.02$. \textbf{c.} HPAM 200 ppm: $\alpha_1 = 4.10^{-4},\,\alpha_2 = -3.10^{-5}$.  \anais{\textbf{d.} Comparison of the integrated pressure from the sensor array $P_{\rm sensors}$ to the mean pressure given by the built-in force sensor of the rheometer $P_{\rm rheometer}$. The dotted line has a slope equal to 1.}}
\end{figure}

\noindent From Figure \ref{figure2}d, the pressure profile $P(r)$ within the sheared polymer solution can be extracted, and plotted for each shear stress $\sigma$. The results are presented in Figure \ref{figure3}a-c, where the pressure $P(r)$ is plotted for varying shear stresses $\sigma$ and polymer concentrations (5000 ppm in Fig. \ref{figure3}a, 1000 ppm in Fig. \ref{figure3}b and 200 ppm in Figure \ref{figure3}c). \anais{The error bars correspond to the error due to the sensor itself (which is done by comparing the pressures of two sensors placed at an equal distance to the center) and from the noise in the pressure measurements.} For all polymer concentrations, the local pressure $P$ is positive, and decreases with the distance $r/R$ to the center of the geometry. This is similar to previous observations of Alcoutlabi \textit{et al.} \cite{Alcoutlabi:2009} in solutions of a different polymer. The pressure profiles $P(r)$ all seem to extrapolate to a positive pressure for $r = R$, which is consistent with a negative value for $N_2$ \cite{Bird:1987, Alcoutlabi:2009}. It should also be noted that the pressures detected here are significantly smaller than the ones measured in other polymer solutions \cite{Alcoutlabi:2009}. In particular, minute pressures (smaller than 10 Pa) that develop in very diluted solutions (figure \ref{figure3}c) are detected, which opens the way to a better characterization of these much less studied systems.

\medskip

\noindent \anais{The experiments are compared with the pressure profile $P(r)$ expected from the theory. Indeed, since our sensors are set to zero at the beginning of the experiment (at zero angular velocity), they are not sensitive to the weight of the sample or to the capillary pressure in the meniscus (if both remain constant as a function of time). Thus, the pressure measured by the sensors directly gives $-\sigma_{zz}$, which is calculated below.} Both gravity and the centrifugal forces are neglected here. 

\noindent From equation \ref{motion}, the equation of motion within the fluid is expressed as a function of $\sigma_{zz}$ and the two normal stresses only \cite{Dbouk:2013}:


\begin{equation}
\label{previous}
    \frac{\partial \sigma_{zz}}{\partial r} = \frac{\partial{N_2}}{\partial r} + \frac{(N_1 + N_2)}{r}.
\end{equation}

\noindent In the HPAM solutions, the normal stresses (figure \ref{figure3}b) are proportional to $\dot{\gamma}^m$, so that $N_1$ and $N_2$ can be written as $N_1 = \alpha_1\, \dot{\gamma}^m$ and $N_2 = \alpha_2 \,\dot{\gamma}^m$, with $\alpha_1$ and $\alpha_2$ the normal stress coefficients. Using these notations, equation \ref{previous} becomes:

\begin{equation}
\label{eqDiff}
        \frac{\partial \sigma_{zz}}{\partial r/R} = \dot{\gamma}^m \left( \alpha_1 + (n+1)\alpha_2 \right)\left(\frac{r}{R}\right)^{m-1}
\end{equation}

\noindent This equation is integrated with respect to $r$, with boundary conditions $\sigma_{rr}(r = R) = -N_2(R)+\sigma_{zz}(R) = P_c$, with $P_c$ the capillary pressure due to the meniscus at the edge of the geometry. Since all sensors are set to zero at the beginning of the experiment (at zero angular velocity), this pre-load is already taken into account and $P_c= 0$. This gives:

\begin{equation}
\label{PProfile}
P(r)= - \sigma_{zz}(r) = - \dot{\gamma}^m \left[ \frac{\alpha_1 + (m+1)\alpha_2}{m}\left(\frac{r}{R}\right)^m - \frac{(\alpha_1 + \alpha_2)}{m}\right].
\end{equation}

\noindent Equation \ref{PProfile} directly gives the pressure profile, which is seen to vary linearly with $(r/R)^m$. This formula is used to determine the two normal stress coefficients $\alpha_1$ and $\alpha_2$. To do so, we combine all experiments at varying $\sigma$ for a given solution by plotting $P/\dot{\gamma}^m$ as a function of $(r/R)^m$. The slope of the linear plot gives  $-\frac{\alpha_1 + (m+1)\alpha_2}{m}$ and the intercept at the origin is $\frac{\alpha_1 + \alpha_2}{m}$, from which $\alpha_1$ and $\alpha_2$ are calculated. The error on $\alpha_1$ is typically of the order of 25\%, while for $\alpha_2$ (which is one order of magnitude smaller than $\alpha_1$) it is of the order of 100\%. The corresponding pressure profiles (with a unique $\alpha_1$ and $\alpha_2$ per concentration) are plotted in Figure \ref{figure3}a-c, with $\alpha_1 = 4.7$ and $\alpha_2 = -0.2$ for $c = 5000$ ppm, $\alpha_1 = 0.46$ and $\alpha_2 = -0.02$ for $c = 1000$ ppm and $\alpha_1 = 4 \times 10^{-4}$ and $\alpha_2 = -3 \times 10^{-5}$ for $c = 200$ ppm. As expected for viscoelastic liquids, $\alpha_1$ is positive and $\alpha_2$ negative, with $|\alpha_1| \gg |\alpha_2|$. Typically, $\alpha_1$ decreases by a factor 10 between the 5000 ppm and the 1000 ppm solutions, and decreases by a factor 1000 when comparing the 1000 ppm to the 200 ppm solutions. Remarkably, we can still measure the normal stress coefficients in the extremely dilute (200 ppm) polymer solution, for which the maximum mean normal stress is smaller than 10 Pa at high shear.
\noindent \anais{To finally check the validity and precision of the sensors, the experimental pressures profiles are integrated to obtain the mean pressure $P_{\rm sensors}$ over the surface of the geometry. In Figure \ref{figure3}d, $P_{\rm sensors}$ (blue disks) is compared to the pressure obtained through the built-in sensor of the rheometer $P_{\rm rheometer}$. The errorbars correspond to the pressure detection threshold (taken at 2 Pa for the sensor array and 4 Pa for the rheometer, see Supplementary Figure 2). All the data collapse on a line with slope 1 (indicated with a dotted line), which confirms that the pressure measured by the sensors is fully consistent with the pressure given by the rheometer instruments.}

\section{\label{sec3}Negative normal stresses}
\begin{figure}[!ht]
\centering
\includegraphics[width=0.65\columnwidth]{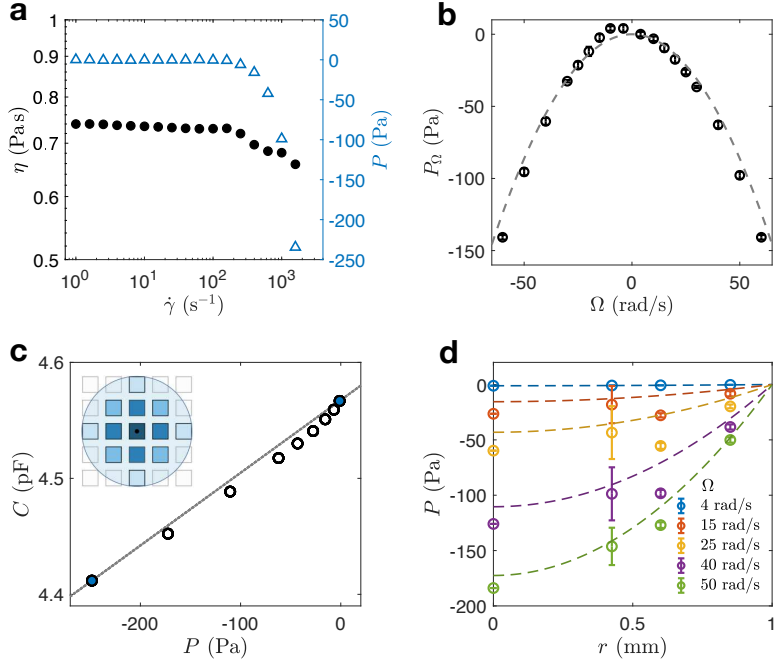}
\caption{\label{figure4}\textbf{a.} Viscosity $\eta$ and mean normal stress (measured by the force sensor of the rheometer) of a 98\% glycerol solution in water as a function of $\dot{\gamma}$. \textbf{b.} Mean normal pressure $P_{\Omega}$ measured through the force sensor of the rheometer, as a function of the rotor angular velocity $\Omega$. The dotted lines are the parabolic fits expected from the theory. \textbf{c.} Calibration curve of a sensor, with (in dotted line) a linear fit. \textbf{d.} Radial distribution of the pressure in a parallel plate torsional flow, for varying angular velocities $\Omega$.
}
\end{figure}

\anais{The sensors are also sensitive to negative normal stresses, which can be measured after a calibration in extension. Here, a negative pressure} is generated by shearing a Newtonian fluid at high angular velocities. We use in this experiment a mixture of 98\% glycerol in water (with density $\rho$ = 1250 kg/m$^3$). \anais{As visible in Figure \ref{figure4}a, the viscosity $\eta$ of the fluid is constant for shear rates $5 \times 10^{-3} < \dot{\gamma} <$ 200 s$^{-1}$ and equal to 740 mPa\,s at 22$^\circ$ C. In a parallel plate geometry as here, the centrifugal forces give rise to a flow circulatory flow visible at high angular velocities $\Omega$, with an inward motion close to the plate and an outward motion near the disk \cite{Savins:1970, McCoy:1971, Turian:1972, Dbouk:2013}, even at relatively low Reynolds number. The impact of this flow becomes visible in our experiment for $\dot{\gamma}>200$ s$^{-1}$ (corresponding to $\Omega >$ 15 rad/s). The recirculation indeed generate negative pressures that are detected by the force sensor: the mean pressure $F_z/\pi R^2$ measured by the rheometer decreases from 0 to -250 Pa while $\dot{\gamma}$ increases from 200 to 1500 s$^{-1}$. In our experiment, the growth of the recirculation is also associated with a slight decrease of the viscosity $\eta$, which cannot be explained by the standard models \cite{Turian:1972}. We interpret this as a destabilisation of the meniscus, leading to the introduction of air in the gap: indeed, the maximum shear rate is close to the limit of liquid ejection  $\dot{\gamma}$ = 3000 s$^{-1}$ in our experiment.}

\noindent \anais{What is to interest for us here is that the recirculation generates a negative pressure profile in the fluid, that can be measured by our sensor arrays. Following \cite{Savins:1970, McCoy:1971, Turian:1972}, the pressure in the flow should} vary quadratically with both the angular velocity $\Omega$ and with the distance $r$ to the center of the geometry:

\begin{equation}
\label{POmega}
\anais{P(r) = -\frac{3}{20}\rho \Omega^2(R^2-r^2).}
\end{equation}

\noindent For a given angular velocity $\Omega$, the average pressure over the surface $S = \pi R^2$ of the geometry is $P_{\Omega} = \frac{1}{\pi R^2}\iint P(r)dS =$ \anais{$-3/40 \rho R^2\Omega^2$}. This quantity is measured directly by the force sensor of the rheometer, which is sensitive to the average pressure over the surface of the geometry. In Figure \ref{figure4}b, the experimental measurement of $P_\Omega$ (circles) is compared to the theory (dotted line), for angular velocities $\Omega$ varying between -60 and 60 rad/s: they are seen to match very convincingly.

\smallskip

We propose to reconstruct the negative pressure profile within the flowing liquid. As presented in the inset of Figure \ref{figure4}c, the sensor array is here centered with respect to the geometry, so that the local pressure $P$ is measured at four different distances $r$ from the center of the geometry. \anais{The sensors are calibrated using the theoretical pressure $P$ (Eq. \ref{POmega}) at the largest angular velocity $\Omega$, and by doing a linear interpolation of the capacitance measured between $\Omega = 0$ and the largest angular velocity (here $\Omega$ = 60 rad/s). Indeed, as presented in Figure \ref{figure4}c, the sensor capacitance varies almost linearly with the pressure $P$ in the region of negative pressures considered here (-300 Pa $<P<$ 0 Pa).} In Figure \ref{figure4}c we show in a dotted line the linear interpolation between the pressures $P = -270$ Pa ($\Omega$ = 60 rad/s) and $P = \Omega = 0$ (blue circles), which models convincingly all other calibration points $C(P)$ (black circles). After calibration, the local pressure profiles $P(r)$ are extracted from the flow. They are presented in Figure \ref{figure4}c, with a different color for each angular velocity $\Omega$ (varied between 4 and 50 rad/s). The pressure profiles $P(r)$ are quadratic, as expected from Equation \ref{POmega}. The theoretical profile is presented with a dotted line: it compares well with the experimental measurements. This demonstrates the potential of the sensors arrays to also detect negative pressures with a reasonable accuracy. In the following, the pressure measurement in an inertial Newtonian fluid flow will be used as a reference flow to calibrate the sensors for negative pressures.

\section{\label{sec4}Evidencing local heterogeneities}

Interestingly, the small size of the sensors \anais{when compared with the size of the geometry}, combined with a high measurement frequency, make them a very good tool to study heterogeneous flows. We consider here a suspension of non-Brownian solid particles (of cornstarch) in a neutrally buoyant Newtonian liquid. At high solid fractions $\phi$, these suspensions exhibit a striking shear-thickening behavior: while they flow easily at low stirring, they become highly viscous at high stirring with an almost solid-like behavior. It is now admitted that this shear-thickening phenomenon associated with an evolution from a lubricated interaction between the particles (at low shear) to a frictional contact (at high shear), due to an increase of the normal stresses that push particles together \cite{Wyart:2014,Seto:2013,Fernandez:2013,Guy:2015,Royer:2016,Clavaud:2017}. Interestingly, close to the shear-thickening transition, large temporal fluctuations in viscosity, shear rate or local density are ubiquitously observed \cite{Lootens:2003,Grob:2016,Hermes:2016}. Recent experiments and new measurement methods (magnetic resonance \cite{Fall:2015}, ultrasound imaging \cite{SaintMichel:2018}, X-ray radiography \cite{Ovarlez:2020} or boundary stress microscopy \cite{Rathee:2017}) seem to indicate that the fluctuations are associated with the presence of local heterogeneities. This observation is further confirmed by recent simulations \cite{Chacko:2018}, but a lot remain to be understood, still, on the exact nature of the heterogeneities and their origin. 

\begin{figure}[!ht]
\centering
\includegraphics[width=0.99\columnwidth]{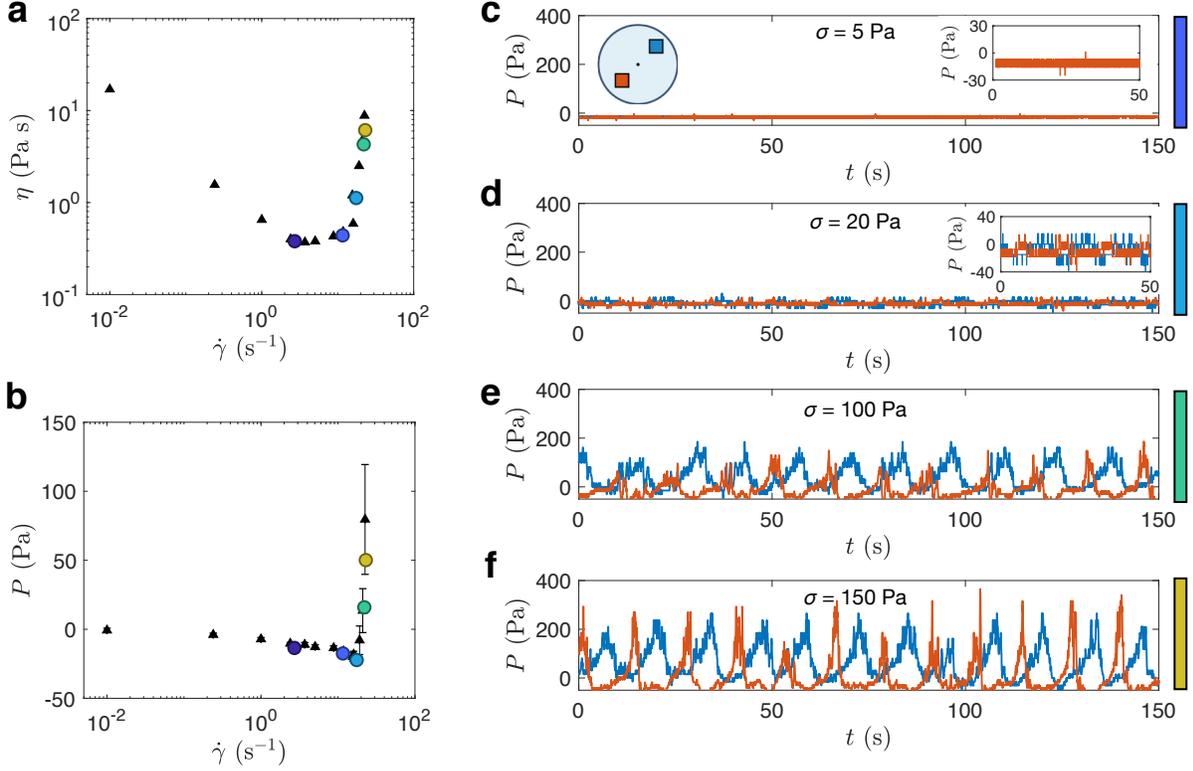}
\caption{\label{figure5}Suspension of 41\% of cornstarch particles in an isodensity water/CsCl solution. \textbf{a.} Viscosity $\eta$ as a function of the shear stress $\dot{\gamma}$. \textbf{b.} Normal stress $P$ measured by the axis of the rheometer as a function of $\dot{\gamma}$. \textbf{c-f.} Normal stress measurement of the sensors as a function of time, for varying imposed constant stresses $\sigma$. Two 4x4 mm$^2$ sensors (in red and blue see inset of \textbf{c}) are placed in opposite directions on a diagonal. \textbf{c.} $\sigma$ = 5 Pa. \textbf{d.} $\sigma$ = 20 Pa. \textbf{e.} $\sigma$ = 100 Pa. \textbf{f.} $\sigma$ = 150 Pa.
}
\end{figure}

\noindent We use a suspension of solid cornstarch particles, a popular shear-thickening system \cite{Fall:2008, SaintMichel:2018}. A large volume fraction of solid particles (here, a 41\% in weight of cornstarch) is put in an isodensity salt of CsCl in water, so that the fluid is in the discontinuous shear-thickening region. As presented in Figure \ref{figure5}a, the 41\% cornstarch suspension is shear-thinning at low shear rate ($\dot{\gamma} < 2$ s$^{-1}$) and strongly shear thickening above a critical shear rate $\dot{\gamma_c} = 5$ s$^{-1}$, with a viscosity $\eta$ increasing by more than one order of magnitude. In figure \ref{figure5}b, we report the mean normal stress generated by the flow as recorded by the rheometer. Here again, two regions can be distinguished: at low shear rates, normal stresses are small and negative and decrease slowly with $\dot{\gamma}$. For a shear rate $\dot{\gamma} > 15$ s$^{-1}$, however, normal stresses become positive and increase rapidly with $\dot{\gamma}$. The error bars indicate the amplitude of the pressure fluctuations, which increase dramatically in the shear-thickening region. 

Our sensor arrays give access to the normal stresses associated with these fluctuations, which has never been done. We use here two sensors, situated at opposite directions below the rotating plate (see the inset of Figure \ref{figure5}a). They record the local pressure at a high frequency of 100 Hz. To study the dynamical behavior of the suspension, we impose 300 s steps of constant shear stresses, either in the shear-thinning region ($\sigma$ = 1 and 5 Pa) or in the shear-thickening region ($\sigma$ = 20, 100 and 150 Pa), points highlighted by colored circles in Figure \ref{figure5}a and b. The local pressures are plotted as a function of time for each stress step in Figures \ref{figure5}c-f. For $\sigma = 5$ Pa, below the discontinuous shear-thickening region, the pressure is constant and identical for both sensors (figure \ref{figure5}c). It is slightly negative ($P \simeq - 10 \pm 3$ Pa, see zoom in the inset), in good agreement with the global pressure measured by the rheometer. However, as soon as the shear thickening region is reached, oscillations become visible. The pressure variations are initially small, of the order of 30 Pa for $\sigma$ = 20 Pa (see Figure \ref{figure5} and inset). They increase in amplitude with increasing $\sigma$: the pressure variation is of the order of 150 Pa for $\sigma$ = 100 Pa and $350$ Pa for $\sigma = 150$ Pa. Very interestingly, the pressure fluctuations are extremely regular, with out of phase peaks of similar amplitude and a constant negative pressure in between. The period $T$ between two peaks slightly decreases with increasing shear stress $\sigma$: $T$ = 15.1 s, 13.3 s and 11.3 s respectively for $\sigma = 20, 100$ and 150 Pa. Interestingly, $T$ is very close to twice the rotational period of the upper disk (respectively equal to 13.8, 11.3 and 10.7 s). All these results seem to indicate the presence of one large aggregate rotating with a angular velocity $\Omega_A \simeq \Omega/2$ within the parallel plate geometry. The presence of this aggregate is associated with localized and very high normal stresses: for $\sigma = 150$ Pa, the peak pressure is 7 times higher than the mean pressure obtained through the force sensor of the rheometer. 
\smallskip

\noindent This observation is in good agreement with the recent work of Ovarlez \textit{et al.}, which evidenced the presence of "density waves" in a similar cornstarch suspension \cite{Ovarlez:2020}. In a Couette cell, the density waves move in the flow direction, more slowly than the rotor, similarly to what is detected here. Our results are also consistent with the observations of Rathee \textit{et al.} \cite{Rathee:2017}, who showed the presence of zones of high shear stress propagating with an angular velocity $\Omega_A = \Omega/2$ in a suspension of sub-micronic silica particles in glycerol. Here, for the first time, we evidence and measure the normal stresses associated with these heterogeneities. The local stresses are 5 to 10 times higher than the average pressure measured by the rheometer, which surely generate an large torque on the geometry. They are thus most likely at the origin of the off-axis motion of the rotor in Couette cells, which was observed in cornstarch \cite{Ovarlez:2020} and in latex particle suspensions \cite{Laun:1994}. It should be noted, finally, that even in a simple parallel plate geometry, the amplitude of the peak pressure remains undetected by the force sensor of the rheometer: the fluctuations are only visible when measured over an area \anais{significantly smaller than the plate size} to avoid a compensation between the low and high pressure regions. 

\section{\label{Conclusion}Conclusion}

\anais{In the study of Newtonian fluids or homogeneous flows, the conventional rheometers, which give access to the average values of the shear stress and normal forces, are particularly relevant. However, heterogeneous flows often occur in complex fluids, as in cornstarch suspensions. The sensor array that we present here is a new tool that brings a new insight into these flows. Here, we access the local value of $\sigma_{zz}$, an important parameter that is much less studied than the shear components of the stress tensor. Our results evidence the potential of the sensor array, not only to measure the normal stresses $N_1$ and $N_2$ (as done with the HPAM solution) but to evidence and follow heterogeneities (as in cornstarch suspensions). It is, to our knowledge, the only sensor capable of evidencing the normal stresses generated by rapidly moving objects, with very high signal to noise  and with a frequency up to 200 Hz. In addition, and in contrast with other local pressure sensors, our system is highly versatile. The size, number and position of the sensors can be modified by replacing the bottom electrode array, which is independent from the measurement surface.\\
\noindent Further developments are also possible: the addition of a soft bottom electrode would make the whole sensor array bendable enough to be added to a Couette geometry. The measurement of normal stresses has never been reported in this very common geometry, despite a huge interest of the community. It would very nicely complement the experiments (of X ray radiography or ultrasound imaging) that have been previously made in such a geometry \cite{Ovarlez:2020, SaintMichel:2018} with heterogeneous flows.}

\bibliographystyle{unsrt}
\bibliography{biblio.bib}

\end{document}


\begin{center}
\large 
\textbf{Pressure sensor arrays for a local normal stress measurement in complex fluids\\}
\vspace{0.3 cm}
\normalsize
\textbf{Supplementary Materials}

\end{center}

\vspace{0.5 cm}

\section*{Supplementary Figures}

\begin{figure}[h!]
\centering
\includegraphics[scale=0.75]{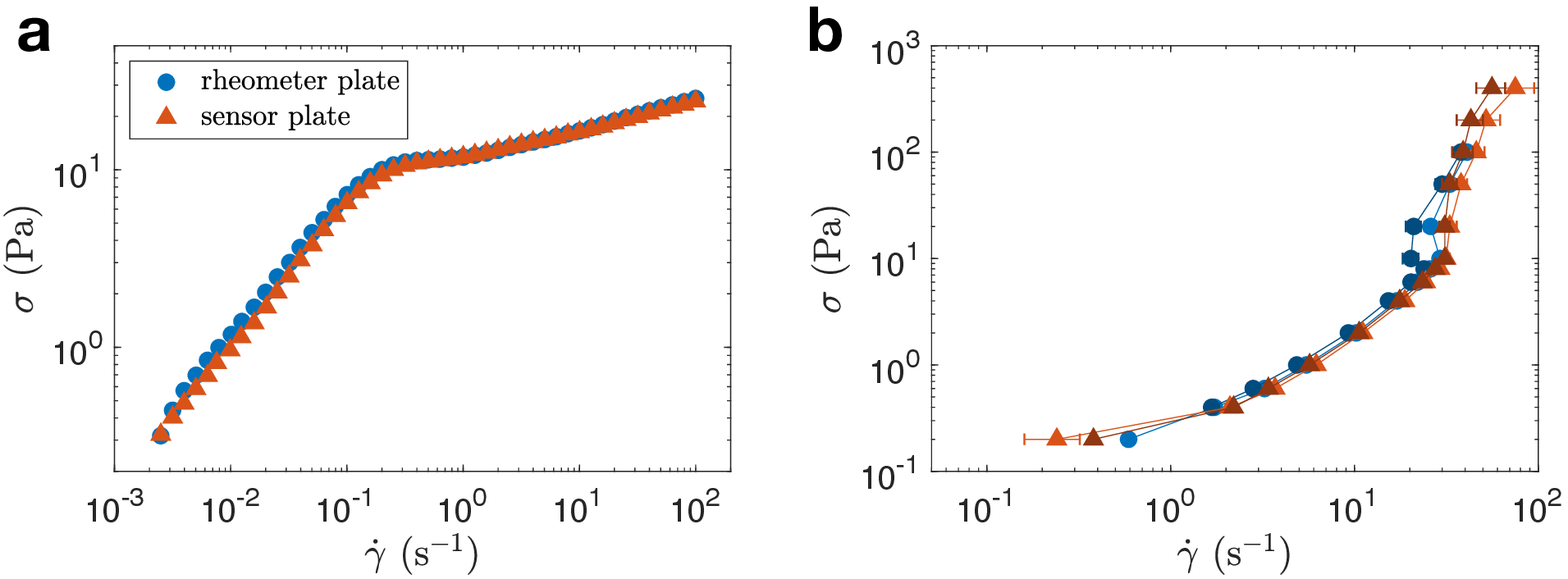}
\caption{\label{SuppFig1} Comparison of the flow curves obtained either on the solid rheometer plate (blue dots) or on the soft sensor array (red triangles). \textbf{a.} Flow curve of a 7000 ppm HPAM solution, with imposed flow sweep from $\dot{\gamma} = 2\times 10^{-3}$ to 100, 10 points per decade. The shear stress is recorded after 10 s of stabilisation.  \textbf{b.} Flow curve of a 40\% cornstarch suspension in an isodensity liquid of water and CsCl salt. Here, the stress $\sigma$ is imposed, with $\sigma$ varied between 0.2 and 400 (4 points per decade). The duration of each steps is 60 s. This shear thickening system naturally exhibits large variations of the flow curve: to evidence this, two experiments are shown, both on the rheometer plate (circles with blue shades) and on the sensor array (triangles with red shades).}
\end{figure}

\begin{figure}[h!]
\centering
\includegraphics[scale=0.7]{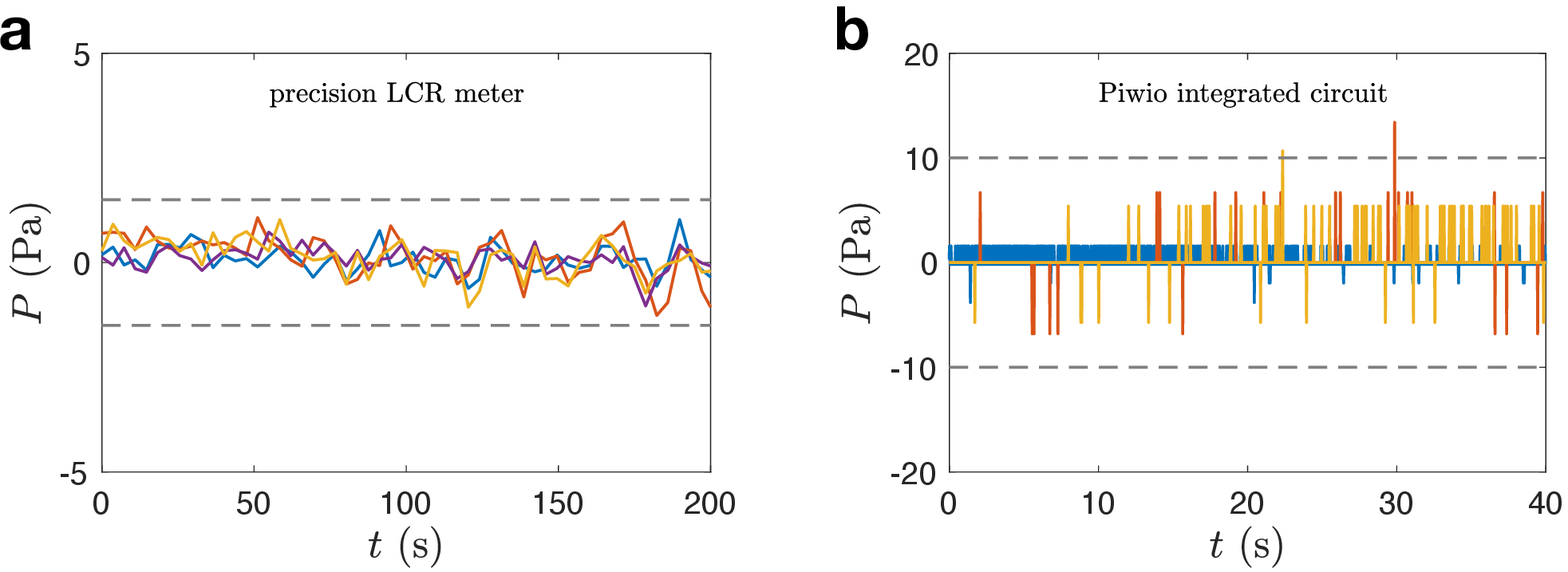}
\caption{\label{SuppFig2} Comparison of the typical response of the sensors (after calibration) as a function of time, using the two different acquisition systems. \textbf{a.} With the precision LCR meter (Agilent technologies), the noise is typically 1.5 Pa, much larger than the resolution of the acquisition system (0.00001 pF, which corresponds to  $10^{-3}$ Pa). \textbf{b.} With the custom-made Piwio board, the measurement is limited by the resolution (of 0.1 pF) and the minimum pressure that can be measured is 10 Pa.}
\end{figure}

\begin{figure}[h!]
\centering
\includegraphics[scale=0.7]{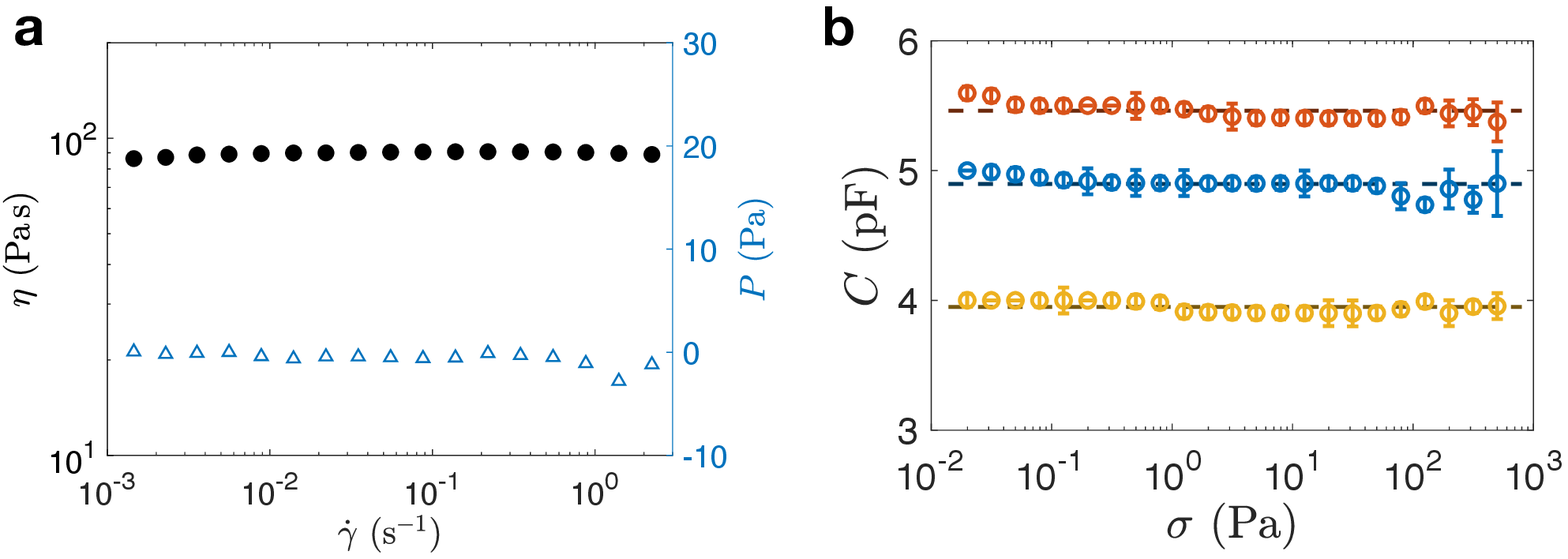}
\caption{\label{SuppFig3} \textbf{a.} Viscosity $\eta$ and mean pressure $P$ measured by the rheometer force sensor for a sheared silicone oil with viscosity 100 000 cSt. \textbf{b.} Response of the sensors in the same experiment as a function of the shear stress $\sigma$. The blue curve corresponds to a sensor centered with respect to the geometry (at $r$ = 0) and the yellow and red curves to two sensors positioned diagonally at a distance $r$ = 12 mm. The capacitance remains constant over 5 orders of magnitude of the shear stress, indicating that they are not sensitive to $\sigma$.}
\end{figure}

\begin{figure}[h!]
\centering
\includegraphics[scale=0.6]{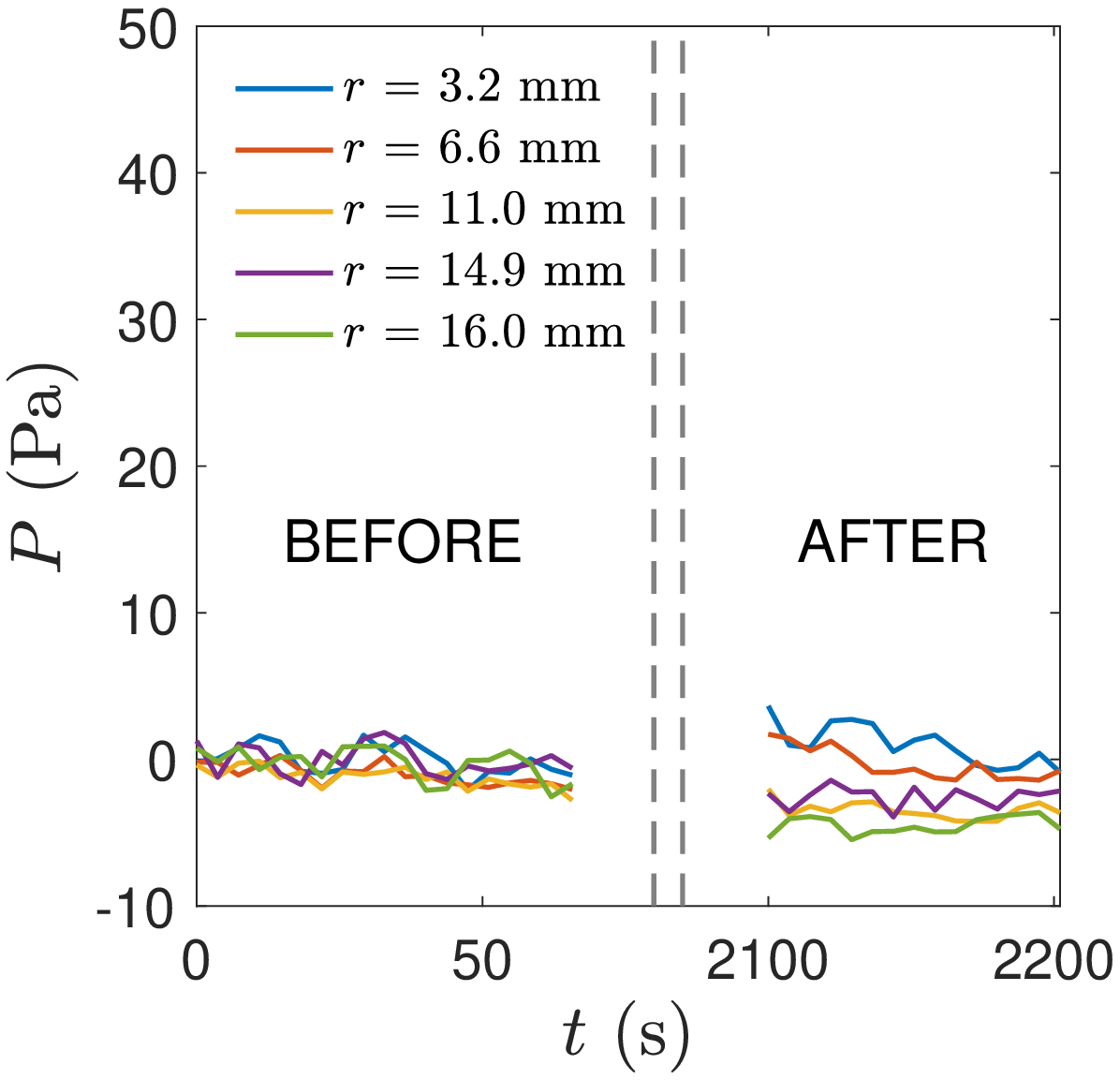}
\caption{\label{SuppFig4} \textbf{a.} Pressure $P$ measured by 5 sensors submitted to a constant hydrostatic pressure of 300 Pa. \textbf{b.} Pressure measured by 5 sensors before and after a 2000 s measurement. The experiment corresponds to the one presented in Figure 2d, with stress-imposed steps of 300 s between $\sigma$ = 0 Pa and $\sigma$ = 20 Pa. While a small hysteresis is observed (the pressure measured by the central sensors is slightly higher than the ones at the edge, corresponding to higher maximal pressures during the experiment), the total scattering of the data is less than 5 Pa. This indicates that there is almost no drift at the timescale of an experiment.}
\end{figure}
\begin{figure}[h!]
\centering
\includegraphics[scale=0.70]{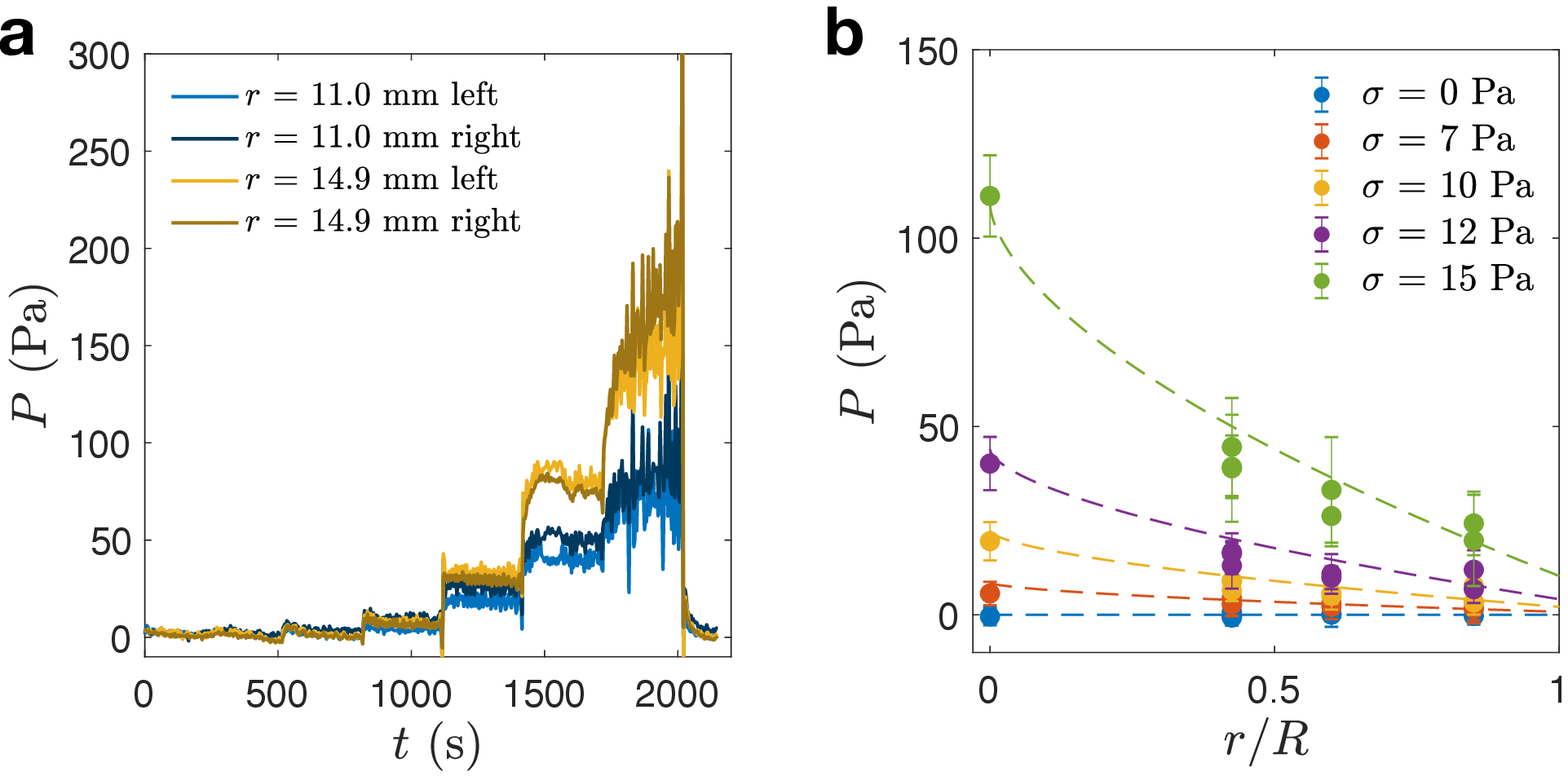}
\caption{\label{SuppFig5} \textbf{a.} Comparison of the pressures $P(t) = -\sigma_{zz}(t)$ in a HPAM solution at 5000 ppm, for increasing shear stresses (the experiment is the one of Figure 2d). \textbf{b.} Pressure profiles measured for a HPAM solution at 5000 ppm for varying shear stresses. The sensor array is centered with respect to the rotating geometry, and each dot corresponds to a different sensor: two sensors placed at an identical distance to the center record very similar pressures. The error bars show the typical noise of the measurement.}
\end{figure}